\documentclass[twocolumn,prl,showpacs,superscriptaddress]{revtex4}
\usepackage{graphicx}
\usepackage{dcolumn}
\usepackage{bm}

\begin{document}

\title{Topological Bose-Mott Insulators in a One-Dimensional Optical
Superlattice}
\author{Shi-Liang Zhu}
\affiliation{National Laboratory of Solid State
Microstructures,Department of Physics,Nanjing University,
Nanjing,China} \affiliation{Laboratory of Quantum Information
Technology and SPTE,South China Normal University,Guangzhou,
China} \affiliation{Department of Physics and Center of
Theoretical and Computational Physics, The University of Hong
Kong,Pokfulam Road, Hong Kong,China}
\author{Z. D. Wang}
 \affiliation{Department of Physics and Center of
Theoretical and Computational Physics, The University of Hong
Kong,Pokfulam Road, Hong Kong,China}
\author{Y.-H. Chan}
\affiliation{Department of Physics, University of Michigan, Ann Arbor, Michigan 48109, USA}
\affiliation{Center for Quantum Information, IIIS, Tsinghua University, Beijing, China}
\author{L. -M. Duan}
 \affiliation{Department of Physics,
University of Michigan, Ann Arbor, Michigan 48109, USA} \affiliation{Center for
Quantum Information, IIIS, Tsinghua University, Beijing, China}

\begin{abstract}
We study topological properties of the Bose-Hubbard model with
repulsive interactions in a one-dimensional optical superlattice.
We find that the Mott insulator states of the single-component
(two-component) Bose-Hubbard model under fractional fillings are
topological insulators characterized by a nonzero charge (or spin)
Chern number with nontrivial edge states. For ultracold atomic
experiments, we show that the topological Chern number can be
detected through measuring the density profiles of the bosonic
atoms in a harmonic trap.
\end{abstract}

\pacs{03.75.Lm, 05.30.Rt, 73.21.Cd}
\maketitle

\textsl{Introduction --} Ultracold atoms in optical lattices can
be used to simulate strongly correlated many-body models that are
central to the understanding of condensed matter physics. This
simulation has attracted a lot of attention as the optical lattice
experiments offer unparalleled
controllability and new tools to study many-body physics \cite%
{Jaksch1998,Greiner2002,Duan2003,Lewenstein2007}. As a remarkable
example, the Bose-Hubbard (BH) model has been experimentally
realized with ultracold atoms and a quantum phase transition from
a superfluid to a Mott insulator described by this model has been
observed \cite{Greiner2002}. On the other hand, topological
matters, such as quantum Hall systems and topological insulators,
are of fundamental importance in physics \cite{Hasan2010}.
Recently, studying topological phases with ultracold atoms has
raised great interest
\cite{Duan2003,Zhu2006,Shao2008,Umucallar2008,Goldman2010,Alba2011,Lang2012,Mei2012}.
In general, it requires complicated control of experimental
systems to realize topological phases with ultracold atoms. An
interesting question is whether one can observe topological phases
in a simple BH type of model, which can be readily implemented by
many experimental groups. Topological properties of bosonic
systems, however, have not been well-studied in literature, partly
for the reason that the topological invariants are usually defined
as an integration over all the occupied states in the momentum
space \cite{Thouless1982,Bohm2003}. This definition does not apply
directly to the bosonic system as many bosons can occupy the same
momentum state.

In this Letter, in contrast to the conventional wisdom, we show
that the BH model in a one-dimensional (1D) optical superlattice
displays nontrivial topological properties. We demonstrate that
the Mott insulators of the single-component (two-component) BH
model at fractional fillings belong to topological matter with its
phase characterized by a nonzero integer charge (or spin) Chern
number. For Mott insulators, the bulk system has a gap in the
excitation spectrum induced by the interaction. For a
topologically nontrivial Mott insulator state characterized by a
nonzero Chern number, we further show that there are protected
edge states inside the bulk gap under the open boundary condition.
The Mott insulators at integer fillings for this system remain
topologically trivial with a zero Chern number and no edge states.
The topological properties discussed here are reminiscent of
those in topological Mott insulators theoretically predicted in Ref. \cite%
{Raghu2008} for the Fermi-Hubbard model in a honeycomb lattice
with frustrated next-neighbor interactions. Remarkably, we  here
show that topological Mott insulators can appear in a simple 1D BH
model in an optical superlattice, which, besides being
conceptually interesting, makes the experimental realization of
topological matter much easier in the ultracold atomic system. We
propose a scheme to detect the topological Chern number by
observation of the plateaus of the density profile with ultracold
atoms in a weak global harmonic trap as it is the case for
experiments.

\textsl{Single-component BH model in a superlattice.--} We consider a
single-component bosonic gas loaded into a 1D optical lattice, which is
described by the BH model
\begin{equation}
H=-J\sum_{\langle ij\rangle }b_{j}^{\dagger }b_{i}+\sum_{j}\left[
Un_{j}(n_{j}-1)/2+V_{j}n_{j}\right] ,  \label{Hs}
\end{equation}%
where $\label{V_j}V_{j}=V\cos (2\pi \alpha j+\delta )$ denotes a
periodic superlattice potential \cite{Roati2008}, $b_{j}$ and
$b_{j}^{\dagger }$ correspond to the bosonic annihilation and
creation operators of atoms on the $j$th lattice site,
$n_{j}=b_{j}^{\dagger }b_{j}$ is the number operator, and $J$ and
$U$ represent the hopping rate and the on site interaction
strength, respectively. We consider in this Letter a commensurate
superlattice potential $V_{j}$ with $\alpha =p/q$ ($p,q$ are
integers) being a rational number and $\delta $ an arbitrary
tunable phase, which has been experimentally realized
\cite{Roati2008}. We take $J$ as the energy unit by setting $J=1$.

The ground-state phase diagram of the Hamiltonian [Eq.(\ref{Hs})]
is well-studied \cite{Rousseau2006,Jaksch1998,Greiner2002}. For a
sufficiently large $U$, the
system is in a gapped Mott insulator phase at commensurate fillings with $%
\nu \equiv N_{b}/N=m\alpha $, where $m$ is an integer, $N_{b}$ is the atom
number, and $N$ is the number of lattice sites. Away from the commensurate
fillings or for a small $U$, the system is in a superfluid state \cite%
{Rousseau2006}. In this Letter, we focus on study of the
topological properties of the Mott insulator phase.

\textsl{The energy gap and the Chern number of the ground state.--} The
topological property is best characterized by the Chern number. To calculate
the Chern number, we first perform exact diagonalization of the Hamiltonian [Eq.(\ref%
{Hs})] on a chain of $N$ sites with periodic or open boundary conditions \cite%
{Zhang2010}. The ground state is nondegenerate and separated from
the higher eigenstates by a finite gap $\Delta $ at the
commensurate fillings.
This gap is shown in Fig.1 as a function of the interaction strength $U$ at $%
\nu =1/3$. The gap increases monotonically with $U$ and then
saturates at a finite value. For a large $U$, the atoms become
hard-core bosons. In this case, each site is occupied by no more
than one atom. The hard-core boson Hubbard model can be mapped to
a model of free fermions. From that mapping, we find that the
saturation value of the gap is $1.08$ (in units of $J$) at a large
$U$ for an infinite
system. The gap should decrease to zero as $U$ drops below a critical value $%
U_{c}$ where the system transits to a superfluid phase. In Fig. 1,
due to the finite size effect, the saturation value of the energy
gap is above $1.08 $ for the large $U$ and does not drop exactly
to zero as $U$ diminishes. However, as the number of lattice sites
increases, we clearly see the tendency that the gap approaches
these limiting values at the two ends.

\begin{figure}[tbph]
\includegraphics[height=4cm]{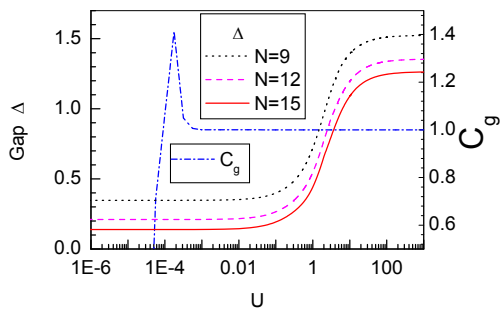}
\caption[Fig. 1 ]{(Color online) The energy gap $\Delta $ and the
Chern number $C_{g}$ [ defined by Eq.(\protect\ref{C_g})] as
functions of the interaction strength $U$. The number of lattice
sites $N=9,12,15$ are used for exact diagonalization, and we take
$N=15$ for calculation
of $C_{g}$. Other parameters include $V=1.5$, $%
\protect\delta =2\protect\pi /3$, $\protect\alpha =1/3$, and $\protect\nu %
=1/3$. }
\end{figure}

Now, we investigate the topological property of the system by
calculating the Chern number. For fermions, the Chern number is
defined as an integration over the occupied states in the momentum
space \cite{Thouless1982}. This definition can not be extended to
the bosonic system as many bosons can occupy the same momentum
state. Fortunately, there is another way to calculate the Chern
number for interacting systems \cite{Niu1985}: suppose the ground
state has a gap to the excited state and it depends on the parameters $%
\theta ,\delta $ through a generalized periodic boundary condition $|\Psi
(j+N,\theta ,\delta )\rangle =e^{i\theta }|\Psi (j,\theta ,\delta )\rangle $,
where $j$ denotes an arbitrary site, $\theta $ is the twist angle, and $%
\delta $ is the phase in the superlattice potential $V_{j}$. Under
this boundary condition, we numerically diagonalize Hamiltonian
[Eq.(\ref{Hs})] and derive the ground state $|\Psi (\theta ,\delta
)\rangle $, which is a non-degenerate state separated from the
excited state by a nonzero energy gap $\Delta $ when $U>U_{c}$.
For the ground state $|\Psi (\theta ,\delta )\rangle $ where
$\theta$ and $\delta $ vary on a torus, one can define the Chern
number $C_{g}$ as a topological invariant by the following formula \cite%
{Niu1985}
\begin{equation}
C_{g}=\frac{1}{2\pi }\int_{0}^{2\pi }d\theta \int_{0}^{2\pi }d\delta
(\partial _{\theta }A_{\delta }-\partial _{\delta }A_{\theta }),  \label{C_g}
\end{equation}%
where the Berry connection $A_{\mu }\equiv i\langle \Psi (\theta
,\delta )|\partial _{\mu }|\Psi (\theta ,\delta )\rangle $ $(\mu
=\delta ,\theta )$. We numerically calculate the Chern number
$C_{g}$ using the method for a discrete manifold \cite{Fukui2005}.
When the parameter $\alpha =1/3$, we find for this boson system
that the Chern number $C_{g}=1$ $(-1)$ for the filling fraction
$\nu =1/3$ $(2/3)$ and $C_{g}=0$ when $\nu =1$. As an example, we
show the value of $C_{g}$ as a function of $U$ at $\nu =1/3$ in
Fig. 1, where the manifold of torus is discretized by $5\times 5$
meshes in the calculation. When the system is in the gapped Mott
state with $U>U_{c}$, $C_{g}$ is quantized to be exactly at $1$,
while $C_{g}$ is unquantized when
the system enters the gapless superfluid phase. Because of quantization of $%
C_{g}$, the finite size effect seems to have a minimal influence,
and we can use exact diagonalization of a small system to get the
exact value of $C_{g}$ in Fig. 1 for the Mott phase (however,
because of the finite-size gap, $C_g$ is still approximately unity
in some region of the superfluid phase near the transition point).
This calculation unambiguously shows that this bosonic system is
in a topological Mott insulator phase with nonzero Chern number at
the fractional filling of the optical lattice.

\textsl{Edge states.-- } The appearance of edge states at the
boundary is usually considered to be a hallmark of nontrivial
topological properties for the bulk system. Under the periodic
boundary condition, this interacting system is gapped at the
fractional filling $\nu =1/3$ $($or $2/3)$. However, under the
open boundary condition, edge states confined to the boundary can
appear inside the energy gap, signaling the nontrivial topological
properties of the bulk insulator. The quasiparticle energy
spectrum $\Delta E_{n}$ is determined by the additional energy
required to add an atom to a system with $n$ atoms, that is,
\begin{equation}
\Delta E_{n}^{(O,P)}\equiv E_{n+1}^{(O,P)}-E_{n}^{(O,P)},  \label{E_ground}
\end{equation}%
where $E_{n}^{(O)}$ ($E_{n}^{(P)}$) is the ground-state energy of the system with $%
n$ atoms under the open (periodic) boundary condition\cite%
{Guo2011}. In Fig. 2(a) , we show the quasi-particle energy spectrum for a
system with $96$ lattice sites near the filling $\nu =1/3$ under both periodic
and open boundary conditions. The calculation is done using the density
matrix renormalization group method \cite{DMRG}, which provides a reliable approach to precisely
calculate energies for any 1D systems. Near the filling $\nu =1/3$, the
quasi-particle energy spectrum is split into two branches separated by a
finite gap. The calculation clearly shows that two states appear in the gap
of the energy spectrum under the open boundary condition. In Fig. 2 (b), we
show the quasi-particle energy spectrum as a function of phase $\delta $
under the open boundary condition. Inside the gap between the lower and the
upper branches of the energy spectrum, one can see two edge modes
(which are degenerate in energy at $\delta =2\pi /3$) that connect these two
branches of the bulk spectrum as one varies phase $\delta $.

To verify that the in-gap modes indeed correspond to the edge states, we
numerically calculate the excitation distribution of these modes and find
that they are confined near the edges of the chain. The distribution of the
quasi-particle can be defined as
\begin{equation}
\Delta n_{j}=\langle \Psi _{n+1}^{g}|n_{j}|\Psi _{n+1}^{g}\rangle -\langle
\Psi _{n}^{g}|n_{j}|\Psi _{n}^{g}\rangle ,  \label{Nj}
\end{equation}%
where $|\Psi _{n}^{g}\rangle $ denotes the ground state wave
function of the system with $n$ bosonic atoms. The distribution of
the in-gap quasi-particle modes for $N=96$ sites under filling
$\nu =1/3$ is plotted in Fig.2 (c). As expected, the in-gap states
mainly distribute near the two edges, especially for a large $V$.
For instance, $99\%$ of the quasi-particle modes at $V=10$ are
localized at the two edge sites.

\begin{figure}[tbph]
\label{Fig2} \includegraphics[width=9cm]{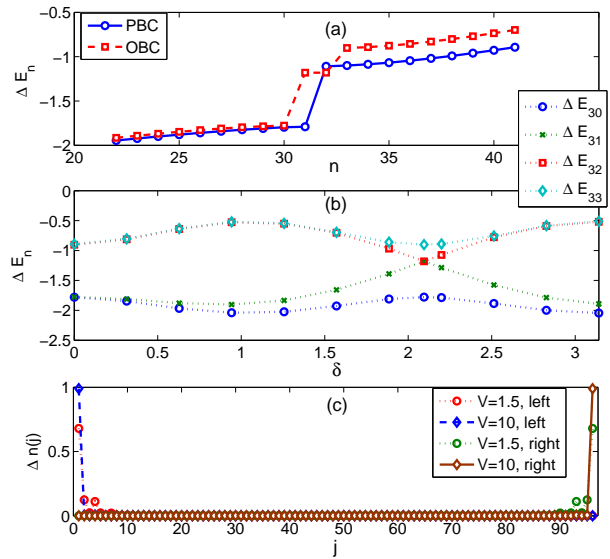} %
\caption{(Color online) (a) The quasi-particle energy spectrum
$\Delta E_n$ [see the definition by Eq. (3)] versus $n$ under the
periodic (PBC) or open (OBC) boundary condition. The calculation
is done in a $96$-site lattice near the filling $\protect\nu=1/3$
with $V=1.5$, $U=10$, $\protect\delta=2\protect\pi/3$, and
$\protect\alpha=1/3$. (b) The edges of the lower ($\Delta E_{30}$)
and the upper  ($\Delta E_{33}$) branches of the energy spectrum
and the two in-gap modes ($\Delta E_{31}$and $\Delta E_{32}$) as
functions of the phase $\delta$ under the open boundary condition.
(c) The distribution of the two in-gap modes along the chain at
$V=1.5,\ 10$. The other parameters for (b) and (c) are the same as
those for (a).}
\end{figure}

\textsl{Two-component BH model in a superlattice and spin Chern
number.--} If the phase $\delta $ in Eq.(2) is replaced by
$-\delta $, we find that the Chern number is $-1$ ($1$) for $\nu
=1/3$ ($2/3$), that is, the sign of the Chern number is flipped.
This fact implies that we can realize a topological insulator
characterized by a nontrivial spin Chern number with a
two-component bosonic gas in a 1D optical superlattice. To this
end, we consider a simple case where the inter-component atomic
collision is turned off, e.g., through Feshbach resonance, and the
system is described by a decoupled two-component BH model with the
Hamiltonian
\begin{equation}
H=-J\sum_{\langle ij\rangle \sigma }b_{i\sigma }^{\dagger
}b_{j,\sigma }+\sum_{j,\sigma }[\frac{U_{\sigma }}{2}n_{j\sigma
}(n_{j\sigma }-1)+V_{j\sigma }n_{j\sigma }],
\end{equation}%
where the potential $V_{j\sigma }=V\cos (2\pi \alpha j+\delta
_{\sigma })$ with $\delta _{\uparrow }=-\delta _{\downarrow
}=\delta $, $b_{j\sigma }$ denotes the bosonic annihilation
operator with the pseudo-spin $\sigma =\uparrow ,\downarrow $, and
$U_{\sigma }$ is the intra-component interaction rate for spin
$\sigma $. For this decoupled two-component BH model, we have the
Chern number $C_{g}^{\uparrow }=-C_{g}^{\downarrow }$. So,
although the total charge Chern number $C_{g}^{\uparrow
}+C_{g}^{\downarrow }$ cancels out to zero, the spin Chern number $%
C_{g}^{s}=C_{g}^{\uparrow }-C_{g}^{\downarrow }$ \cite{Sheng2006}
is non-vanishing at fractional fillings. The nonzero spin Chern
number is usually associated with the quantum spin Hall effects in
two-dimensional systems \cite{Sheng2006}. For our 1D system, spin
edge states appear when the spin Chern number is nonzero. For an
example with $\alpha =1/3$, we
have $C_{g}^{s}=2$ at the fractional fillings $\nu =1/3$ ($2/3$) and $%
C_{g}^{s}=0$ at the integer filling. The edge states are similar to those
shown in Fig. 2. The spin up (down) edge state is confined near the left
(right) edge, respectively.

\textsl{Experimental detection.--} We now discuss how to measure topological
Chern number in a practical experimental setting. For atomic experiments,
apart from the optical superlattice potential, the bosons are confined in a
weak global harmonic trap. For simplicity, we consider the large-$U$ limit
where the system is described by hard-core bosons with no more than one atom
occupying the same lattice site. The total potential, including the optical
superlattice and the global harmonic trap, is described by
\begin{equation}
V_{j}=V\cos (2\pi \alpha j+\delta )+V_{H}(j-j_{0})^{2},
\end{equation}%
where $j_{0}$ denotes the position of the trap center and $V_{H}$ is the
strength of the harmonic trap. We use the Jordan-Wigner transformation, $%
b_{j}^{\dagger }=f_{j}^{\dagger }\prod_{m=1}^{j-1}e^{-i\pi f_{m}^{\dagger
}f_{m}}$ and $b_{j}=\prod_{m=1}^{j-1}e^{i\pi f_{m}^{\dagger }f_{m}}f_{j}$,
to map the hard-core BH\ model to non-interacting fermion Hamiltonian $%
H_{F}=-J\sum_{j}(f_{j}^{\dagger }f_{j+1}+h.c.)+\sum_{j}V_{j}f_{j}^{\dagger
}f_{j}$, where $f_{j}^{\dagger }$ and $f_{j}$ are the creation and
annihilation operators for spinless fermions, respectively\cite{Rousseau2006,Paredes2004}%
. The particle density of hard-core bosons coincides with that of
non-interacting fermions as we have $n_{j}=\langle b_{j}^{\dagger
}b_{j}\rangle =\langle f_{j}^{\dagger }f_{j}\rangle =n_{j}^{F}$
with the Jordan-Wigner transformation; however, the momentum
distribution for bosons is typically very different from that for
fermions.

After the hard-core bosons are mapped to fermions, there is a
simple way to figure out the Chern number. The ground state of
free fermions is a Slater determinant, i.e., a product of single
particle states $|\Psi _{g}^{F}\rangle
=\prod_{m=1}^{N_{f}}\sum_{n=1}^{N}P_{nm}f_{n}^{\dagger }|0\rangle
$, with $N_{f}=N_{b}$ the number of fermions and $P$ the matrix of
the components of $|\Psi _{g}^{F}\rangle $. Supposing that the
$n$-th eigenstate of a single particle is denoted by $|\psi
_{n}\rangle
=\sum_{j}\phi _{j,n}f_{j}^{\dagger }|0\rangle $, the eigenvalue equation $%
H_{F}|\psi _{n}\rangle =E_{n}|\psi _{n}\rangle $ can be written in
terms of the following Harper equation \cite{Lang2012}
\begin{equation}
-J(\phi _{j+1,n}+\phi _{j,n})+V\cos (2\pi \alpha +\delta )\phi
_{j,n}=E_{n}\phi _{j,n},  \label{Harper}
\end{equation}%
where $\phi _{j,n}$ is the amplitude of the particle wave function of the $j$%
-th site and $E_{n}$ is the $n$-th single-particle eigen-energy.
Compared with the Harper equation in a magnetic field, we know
that $\alpha $ corresponds to the magnetic flux \cite{Bohm2003}.
Therefore, we can define the local density difference as
\begin{equation}
\delta n_{j}=\frac{n_{j}(\alpha _{1})-n_{j}(\alpha _{2})}{\alpha _{1}-\alpha
_{2}}.  \label{Streda}
\end{equation}%
The Chern number $C_{g}$ can then be obtained through the Streda formula $%
C_{g}=\delta n_{j}$ under the condition that $n_{j}(\alpha _{\eta })$ $(\eta
=1,2)$ is the local density associated with the plateau at $\alpha _{\eta }$%
\cite{Umucallar2008,Shao2008,Lang2012,Mei2012}.

\begin{figure}[tbph]
\includegraphics[height=4.5cm]{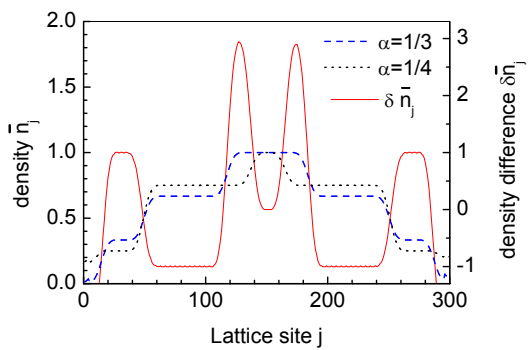} 
\caption{(Color online) The average density profiles $\bar{n}_j$ and the
density difference $\protect\delta \bar{n}_j$ for $\protect\alpha_1=1/3$ and
$\protect\alpha_2=1/4$. The values of $\protect\delta \bar{n}_j$ represent
the corresponding Chern numbers at the plateaus with the fillings $\protect%
\nu=1/3,2/3,1$ for $\protect\alpha=1/3$ and $\protect\nu=1/4, 3/4,1$ for $%
\protect\alpha=1/4$. The other parameters are $N=300$, $N_b=180$,
$M=4$, $V=10$, $\delta=\pi/2$, and $V_H=0.001$. }
\label{Fig3}
\end{figure}

Following the method outlined in Ref. \cite{Rousseau2006}, we
numerically calculated the average density profiles for $\alpha
=1/3,1/4$, with the results shown in Fig.3. To reduce the
oscillations in density profiles induced by modulation of the
potentials, we define the local average density
$\bar{n}_{j}=\sum_{m=-M}^{M}n_{j+m}/(2M+1)$, where $2M+1$ is the
length to average the density, which corresponds to the position
resolution in the experimental detection. We take $M\ll N$, e.g.,
$M=4$ and $N=300$\ in Fig. 3, as it is typical for experiments. \
As one can see from the density
profiles $\bar{n}_{j}$ in Fig. 3, plateaus appear at the rational fillings $%
\nu =1/3,2/3,1$ for $\alpha =1/3$, and $\nu =1/4,3/4,1$ for
$\alpha =1/4$ (the gap at half filling in the case of $\alpha
=p/q$ with an even $q$ is generally closed at an integer $\delta
/\pi $ \cite{Bohm2003}). Using the Streda formula [Eq.
(\ref{Streda})], we obtain $C_{g}=\delta n_{j}=1,-1$ at the
fractional fillings $\nu =\alpha ,1-\alpha $, and $C_{g}=0$ at the
integer filling $\nu =1$. The width of the plateaus is associated
with the size of the energy gap. To make detection of the Chern
number easier, we can adjust the frequency of the harmonic trap to
move the target plateaus to the center of the trap. For example,
if we choose $V_{H}=10^{-4}J$ and other parameters
as those given in Fig.3, the plateaus at $\nu =2/3$ for $\alpha =1/3$ and $%
\nu =3/4$ for $\alpha =1/4$ are moved to the center of the trap spanning
from the $65$th to the $235$th lattice site. With such a wide
plateau, it is straightforward to read out the Chern number $C_{g}=\delta \bar{n}%
_{j}=-1$ for this case.

In summary, we have shown that for bosonic atoms in a 1D optical
super-lattice, the Mott insulator states of the corresponding BH
model at fractional fillings are topologically nontrivial,
characterized by nonzero Chern number and existence of edge
states. We further predict that the topological Chern number can
be detected by measuring the plateaus in the density profile when
the atoms are trapped in a global harmonic potential. The model
discussed in this Letter represents one of the simplest
experimental systems to show intriguing topological properties,
and the proposed detection method allows one to confirm these
topological properties with the state-of-the-art technology.

SLZ is supported by the NSF of China (Grant No. 11125417), the
SKPBR of China (Grant No.2011CB922104), and the PCSIRT. ZDW is
supported by the GRF (HKU7058/11P) and the CRF (HKU8/11G) of Hong
Kong RGC. LMD and YHC acknowledge support by the NBRPC (Grant
No.2011CBA00302), the DARPA OLE Program under ARO Award
W911NF0710576, the IARPA, and the AFOSR and ARO MURI program.

\end{document}